\documentclass[reprint,amsmath,amssymb,aps,showpacs,superscriptaddress,twocolumn,longbibliography,prb]{revtex4-1}
\usepackage{graphicx}
\usepackage{bm}
\usepackage{epsfig}
\usepackage{amssymb}
\usepackage{amsfonts}
\usepackage{braket}
\usepackage{color}
\usepackage{epstopdf}
\usepackage{hyperref}
\usepackage{float}
\restylefloat{table}
\usepackage{bibentry}
\usepackage{multirow}
\usepackage[caption=false]{subfig}

\graphicspath{{figures/}}% Put all figures in this directory.

\begin{document}
\title{ Abelian and Non-Abelian Chiral Spin Liquids in a Compact Tensor Network Representation }

\author{Hyun-Yong Lee}
\email{hyunyong@issp.u-tokyo.ac.jp}
\affiliation{Institute for Solid State Physics, University of Tokyo, Kashiwa, Chiba 277-8581, Japan}

\author{Ryui Kaneko}
\email{rkaneko@issp.u-tokyo.ac.jp}
\affiliation{Institute for Solid State Physics, University of Tokyo, Kashiwa, Chiba 277-8581, Japan}

\author{Tsuyoshi Okubo}
\email{t-okubo@phys.s.u-tokyo.ac.jp}
\affiliation{Department of Physics, University of Tokyo, Tokyo 113-0033, Japan}

\author{Naoki Kawashima}
\email{kawashima@issp.u-tokyo.ac.jp}
\affiliation{Institute for Solid State Physics, University of Tokyo, Kashiwa, Chiba 277-8581, Japan}

\date{\today}

\begin{abstract}
	We provide new insights into the Abelian and non-Abelian chiral Kitaev spin liquids on the star lattice using the recently proposed loop gas\,(LG) and string gas\,(SG) states\,[H.-Y. Lee, R. Kaneko, T. Okubo, N. Kawashima, Phys. Rev. Lett. 123, 087203 (2019)]. Those are compactly represented in the language of tensor network. By optimizing only one or two variational parameters, accurate ansatze are found in the whole phase diagram of the Kitaev model on the star lattice. 
	In particular, the variational energy of the LG state becomes exact\,(within machine precision) at two limits in the model, and the criticality at one of those is analytically derived from the LG feature. 
	It reveals that the Abelian CSLs are well demonstrated by the short-ranged LG  while the non-Abelian CSLs are adiabatically connected to the critical LG where the macroscopic loops appear. Furthermore, by constructing the minimally entangled states and exploiting their entanglement spectrum and entropy, we identify the nature of anyons and the chiral edge modes in the non-Abelian phase with the Ising conformal field theory. 
\end{abstract}
\maketitle

\section{introduction}

Discovery of the fractional quantum Hall\,(FQH) effect\cite{Tsui82} had brought a paradigm shift in understanding of condensed phases of matter. Exotic quantum liquid states, i.e., chiral spin liquids\,(CSL), were proposed as the ground states of the FQH system\cite{Laughlin87,Wen89}, which cannot be featured by Landau's symmetry breaking theory but the so-called {\it topological order}. The topological order can be interpreted as the pattern of long-range entanglement which leads to the ground state degeneracy depending only on the topology of system\cite{Wen90}. Those CSL ansatze successfully explained the nature of the FQH fluids such as the fractional statistics of quasiparticles\,(or anyons)\cite{Arovas84,Halperin84}. Furthermore, the anyons obeying the non-Abelian braiding statistics were theoretically realized in the FQH system\cite{Wen91, Moore91,Ronny09}. Due to the robust topological degeneracy against the local perturbations and exotic statistics of anyons, the non-Abelian topological states have been proposed as a promising platform for fault-tolerant quantum computing\cite{Kitaev2003} and thus attracted lots of attention in the field of quantum information for the last decade\cite{Nayak2008}. Another interesting feature of the FQH fluids and CSLs is that the chiral gapless edge modes appear at the boundary of the system, and it leads to perfect heat conduction at the edge\cite{Wen91a}. The edge states are described by the conformal field theories\,(CFT) which also characterize and hence have been employed to identify the topological order\cite{Li08, Qi12, Zhu15, Shuo15, Poiblanc15, Poiblanc16, Poiblanc17, Chen18}. By solving the Kitaev model\cite{Kitaev2006} on the star lattice\,(KSM), Yao and Kivelson showed the existence of the CSL as an exact ground state of local Hamiltonian and found Abelian and non-Abelian phases characterized by the topological degeneracies four and three on the torus, respectively\cite{Yao07,Chung10}.

The aim of this Letter is to understand the Abelian and non-Abelian Kitaev CSLs without referring to the Majorana fermion. Recently, a particular LG state and its extension, which is referred to as SG state, have been proposed as ansatze for the Kitaev spin liquid\,(KSL) on the honeycomb lattice\cite{Kitaev2006} in a compact tensor product state\,(TPS) representation\cite{HY19}. The LG ansatz was found to reflect most qualitative features of the KSL, and SG provides a quantitatively accurate approximation to the KSL while keeping the qualitative features intact. In what follows, we reinterpret the CSLs as the LG and SG states and provide direct evidences identifying the topological order in each phase.

\begin{figure}[!t]
	\includegraphics[width=0.49\textwidth]{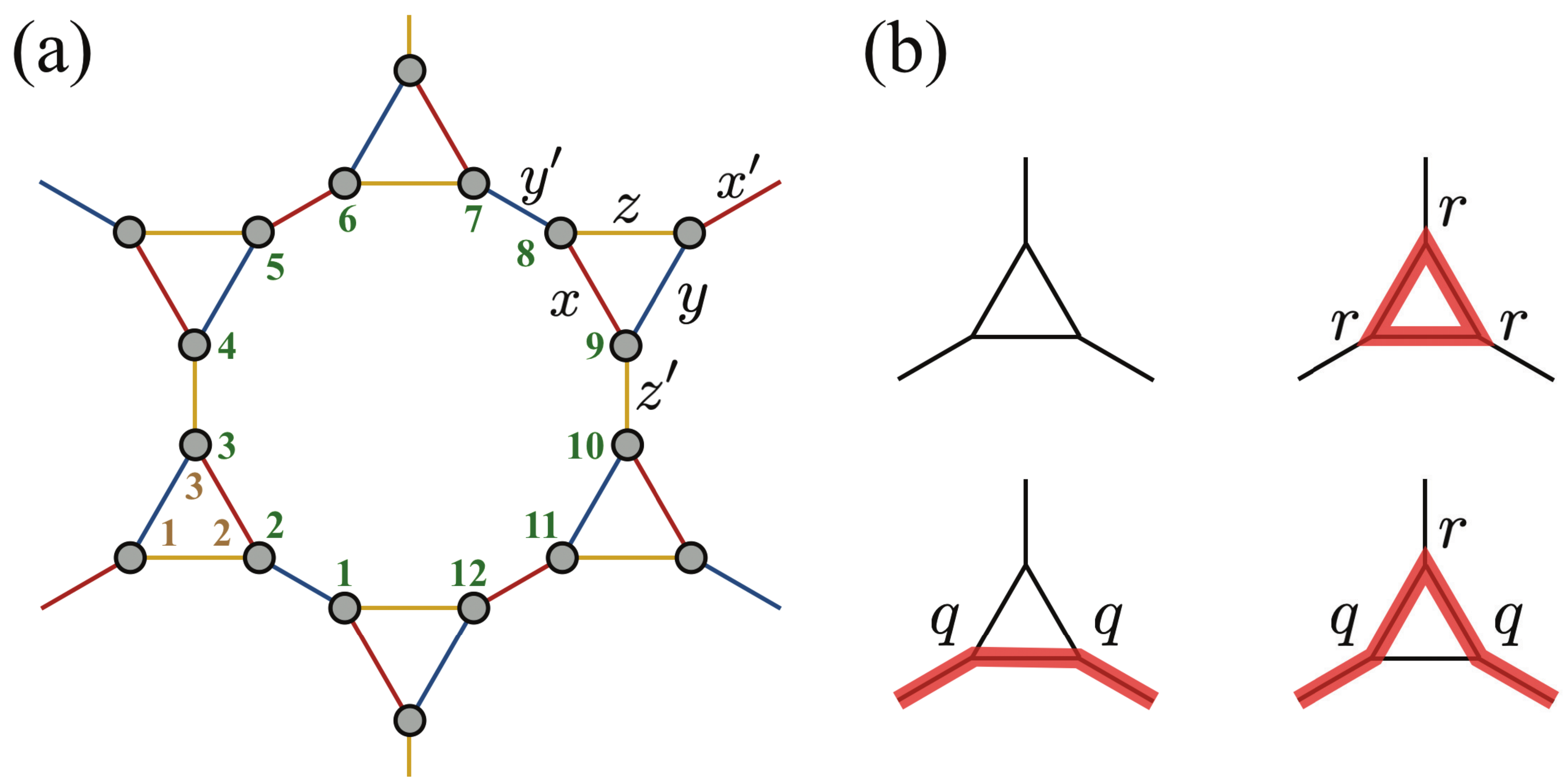}
	\caption{ (a) Schematic figures of the star lattice where the $x$-, $y$- and $z$-bonds defined in the model\,[Eq.\,\eqref{eq:hamiltonian}] are specified by red, blue and yellow colors, respectively.
	(b) Four local configurations and corresponding magnetic states of the LG state\,[Eq.\,\eqref{eq:lg_state}] where $|\theta,\gamma\rangle$ is defined in Eq.\,\eqref{eq:d2_inital_state}, and $x,y,z$ denote each bond of the model.
	    }
	\label{fig:schematic}
\end{figure}
\section{model}

The KSM is defined as\cite{Yao07}
\begin{align}
	\hat{\mathcal{H}} = -\frac{J}{4}\sum_{\langle ij \rangle \in{\gamma }} \hat{\sigma}_i^\gamma \hat{\sigma}_j^\gamma
	-\frac{J'}{4}\sum_{\langle ij \rangle \in{\gamma'}} \hat{\sigma}_i^{\gamma'} \hat{\sigma}_j^{\gamma'},
	\label{eq:hamiltonian}
\end{align}
where $\hat{\sigma}_i^\gamma$ stands for the Pauli matrix with $\gamma,\gamma' = x,y,z$, while $\langle ij \rangle_{\gamma}$ and $\langle ij \rangle_{\gamma'}$ denote the nearest-neighbor pair respectively on the intra-triangle\,($\gamma$) and inter-triangle\,($\gamma'$) bonds connecting sites $i$ and $j$ as defined in Fig.\,\ref{fig:schematic}\,(a). Note that the Hamiltonian commutes with two types of flux operators defined on the triangle plaquette $\hat{V}_p = \hat{\sigma}_1^x \hat{\sigma}_2^y \hat{\sigma}_3^z$ and dodecagon plaquette $\hat{W}_p = \hat{\sigma}_1^x \hat{\sigma}_2^z \hat{\sigma}_3^y \hat{\sigma}_4^x \cdots \hat{\sigma}_{12}^y$\cite{Yao07}, where the site indices are defined in Fig.\,\ref{fig:schematic}\,(a). Therefore, the Hamiltonian is block-diagonalized, and each block is characterized by the set of the flux numbers or eigenvalues of flux operators $\{V_p=\pm1,W_p=\pm1\}$. Since the operator $\hat{V}_p$ consists of three Pauli matrices, the time-reversal transformation flips its flux number, i.e., $\mathcal{T} \hat{V}_p \mathcal{T}^{-1} = -\hat{V}_p $, and hence the TR-symmetry is spontaneously broken in eigenstates of $\hat{\mathcal{H}}$. It was found\cite{Yao07} that the ground states do not break any lattice symmetry\,(that is, the CSLs) and live in the vortex-free sector, i.e., $\{\hat{W}_p=1,\hat{V}_p=1\}$. In addition, the model exhibits a topological phase transition between the non-Abelian and Abelian CSLs at $J'/J=\sqrt{3}$\cite{Yao07}.

\section{variational ansatze}
\subsection{Loop gas states}

\begin{figure}[!t]
	\includegraphics[width=0.5\textwidth]{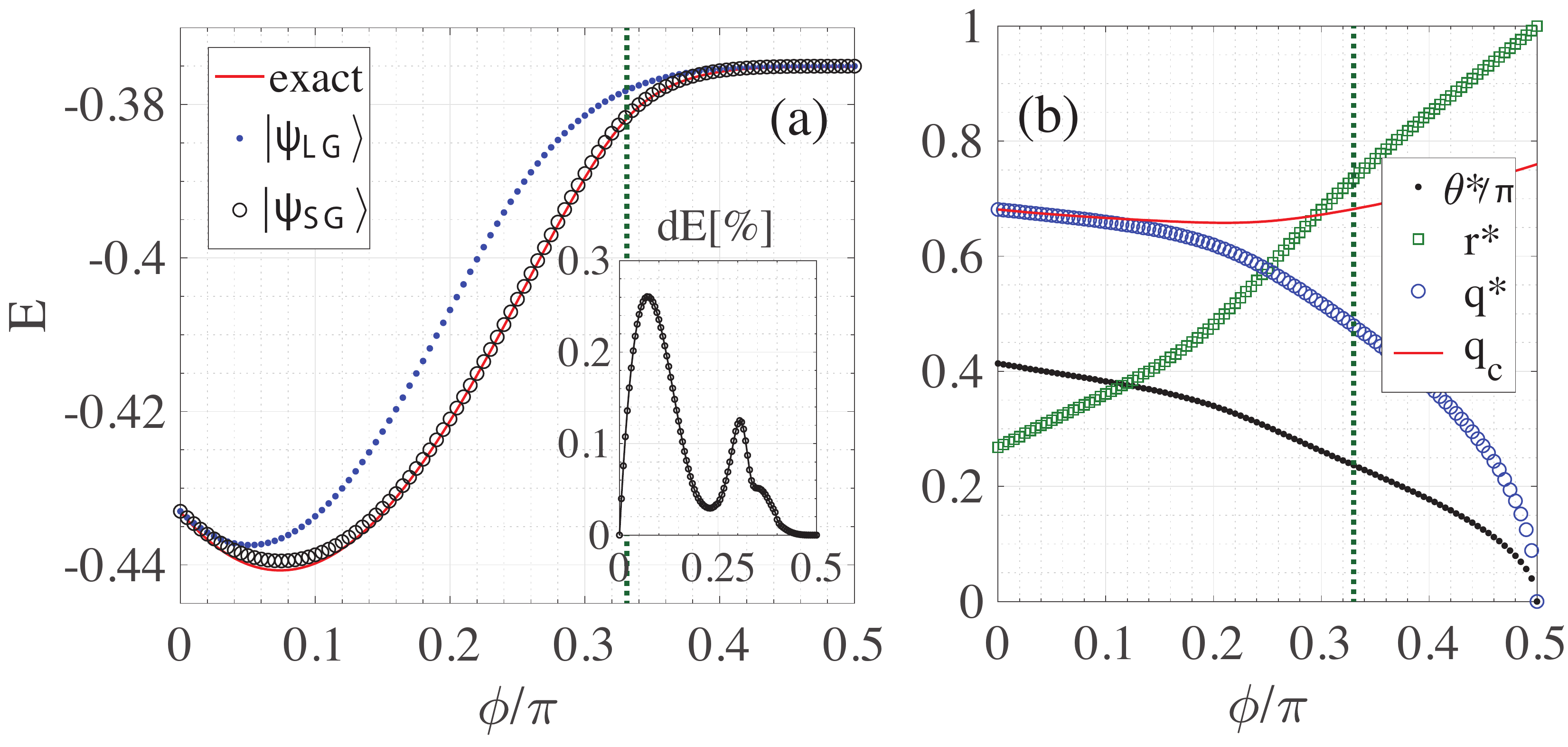}
	\caption{ (a) Variational energy of the optimized LG and SG ansatze\,(inset: $d E=1-E/E_{\rm exact}$ of SG)   
	(b) Optimal variational parameter $\theta^*$ of $|\psi_{\rm LG}\rangle$ as a function of $\phi$. Here, $r^*$ and $q^*$, which are determined by $\theta^*$, are the optimal local weights of loop along the triangle and dodecagon plaquettes, respectively.}
	%The weight $q_c$ indicates the critical weight for a given $r^*$ where the LG state becomes critical. The topological phase transition point is denoted by the green dotted line. 
	\label{fig:energy}
\end{figure}

We begin with defining a product state $|\Psi(\theta)\rangle = \otimes_{\alpha} |\theta,\gamma_\alpha \rangle_\alpha$ with a local magnetic state given by
\begin{align}
	\langle \theta,\gamma | \hat{\sigma}^{\gamma'} |\theta,\gamma \rangle = \delta_{\gamma \gamma'} \cos\theta + (1-\delta_{\gamma \gamma'}) \frac{\sin\theta}{\sqrt{2}},
	\label{eq:d2_inital_state}
\end{align}
and $\gamma_\alpha$ being the {\it inter-triangle} bond at site $\alpha$. For instance, at site-8 in Fig.\,\ref{fig:schematic}\,(a), we assign the state $|\theta, y\rangle$. Applying the loop gas\,(LG) operator $\hat{Q}_{\rm LG}$ defined in Ref.\,\cite{HY19} on top of $|\Psi(\theta)\rangle$ results in a LG state $|\psi_{\rm LG}(\theta)\rangle \equiv \hat{Q}_{\rm LG}|\Psi(\theta)\rangle$, which is simply a superposition of all possible loop configurations of magnetic states, i.e.,
\begin{align}
  \includegraphics[width=0.48\textwidth]{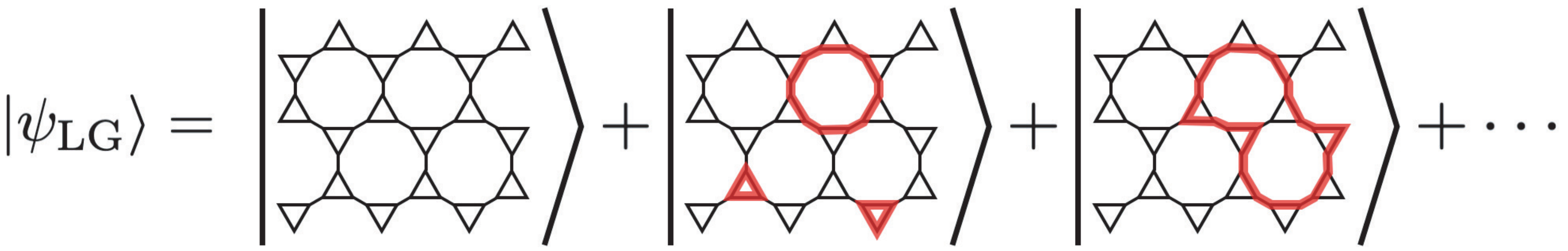}.
  \label{eq:lg_state}
\end{align}
Here, the empty site denotes the state $|\theta,\gamma_\alpha\rangle$ while the loop-occupied site stands for $\hat{\sigma}^\gamma|\theta,\gamma_\alpha\rangle$ depending on the direction of loop as depicted in Fig.\,\ref{fig:schematic}\,(b). 
Due to the LG operator, the desired symmetries, $Z_2$ gauge redundancy and the vortex-freeness are guaranteed in the LG state\cite{HY19}. Notice that the norm of $|\psi_{\rm LG}(\theta)\rangle$ maps into the partition function of the classical $O(1)$ LG model\cite{Nienhuis1982} with the local weights of loops being $r = \cos\theta$ and $q=\sin\theta/\sqrt{2}$ along the triangle and dodecagon plaquttes, respectively, on the star lattice\cite{HY19}. After simple algebra, the partition function can even be mapped into that on the honeycomb lattice, i.e., $Z_{O(1)}(x)$ with $x$ being the fugacity per site:
\begin{align}
	\langle \psi_{\rm LG}(\theta) |\psi_{\rm LG}(\theta)\rangle 
	%& = N_{\Gamma'} \sum_{G\in \Gamma} [q^2(1+r)]^{l_G} (1+r^3)^{2N-l_G}\nonumber\\
	 = c\, Z_{O(1)} \left( \frac{q^2}{1-r+r^2} \right),
	\label{eq:o1_classical}
\end{align}
where $c$ is a constant(see Appendix\,\ref{app:norm}). Now, one can optimize the variational parameter $\theta$ to minimize the energy for a given $(J,J')$. For simplicity, let us parameterize the exchange couplings as $J'/J = \tan\phi$. 
%Then, the topological phase transition occurs at $\phi = \pi/3$, below which the non-Abelian phase appears. 
However, at $\phi = 0$, we consider $J\rightarrow \infty$ while $J'=1$ being finite to avoid the trivial solution at $J'=0$, and vice versa at $\phi=\pi/2$. We employ the corner transfer matrix renormalization group\,(CTMRG) method\cite{Nishino1996,Orus2009,Corboz2010} to measure the energy $E=\langle \psi_{\rm LG} | \hat{\mathcal{H}} | \psi_{\rm LG}\rangle$ and find $\theta^*(\phi)$ minimizing the energy at a given $\phi$. The resulting $E$ and $\theta^*$ are presented in Fig.\,\ref{fig:energy}\,(a) and (b), respectively, as a function of $\phi$. Here, we present the exact energy $E_{\rm ex}$.
% together, which is obtained by integrating the Majorana fermion bands. 
The optimal local weights $(q^*,r^*) = (\sin\theta^*/\sqrt{2},\cos\theta^*)$ are also presented in Fig.\,\ref{fig:energy}\,(b), which provide new insights into the nature of each phase. As one can see, the variational energy of the LG ansatz is quite accurate in $0.4\pi < \phi \leq 0.5\pi$, where the energy deviation $d E=1-E/E_{\rm exact}$ is less than $0.1\%$ as shown in the inset of Fig.\,\ref{fig:energy}\,(a). In particular, the energy becomes exact (within machine precision) at $\phi=0.5\pi$ at which $r$ is maximized while $q$ vanishes as shown in Fig.\,\ref{fig:energy}\,(b). It indicates that the configurations with only triangle loops and holes survive. In this sense, the Abelian CSL phase can be understood as the triangle loop gas in which longer loops are suppressed. 

Interestingly, the variational energy becomes also exact\,(within machine precision) as approaching the opposite limit $\phi=0$ where the excitation gap closes as $J'^2/J$\cite{Yao07, Vidal08}. Notice that $Z_{O(1)}(x)$ in Eq.\,\eqref{eq:o1_classical} becomes critical at $x_c=1/\sqrt{3}$\cite{Nienhuis1982}, which leads to the {\it critical} LG state at $(q_c, r_c) = (\sqrt{2\sqrt{3}-3}, 2-\sqrt{3})$. Remarkably, the LG state is optimized with $(q_c, r_c)$ at $\phi= 0$ which implies that the ground state is the critical LG state exhibiting macroscopic loops. Furthermore, its low-energy physics is described by the Ising CFT\cite{Nienhuis1982} which is consistent with the expected one\cite{Yao07, Teresia13, Lahtinen14, Meichanetzidis16}. Therefore, according to these circumstantial evidences, we may conclude that the LG ansatz at $\phi=0$ is the exact ground state. 
It tells us that the non-Abelian CSL around $\phi=0$ are well described by the long-ranged LG states which is qualitatively distinct from the short-ranged feature of the Abelian CSL. Excluing $\phi=0$, all LG ansatze map into the gapped phase of $Z_{O(1)}(x)$, i.e., $x<1/\sqrt{3}$, ensuring its gapped nature.

%We believe that such fate of the long-ranged loops at each limit is closely related with the non-Abelianess and Abelianess of CSL. 
%It is also worth noting that the optimal initial state at $\phi=0.25\pi$\,($J=J'$) is $|\Phi_{(111)}\rangle = |\Phi(\sin^{-1}\sqrt{2/3})\rangle$ as naively expected. 

%Although the LG ansatz provides exact representation of ground states at both limits and accurate variational energies in the Abelian phase, it neither represents the non-Abelian CSLs very well nor captures the phase transition to the Abelian phase. there  

%
\subsection{String gas states}

In the non-Abelian phase and around the phase transition point, the variational energies of the LG ansatz are away from the exact ones. Therefore, in order to lower the energy and find better ansatze, we apply the dimer gas\,(DG) operator $\hat{R}_{\rm DG}$ as suggested in Ref.\,\cite{HY19},
\begin{align}
  \includegraphics[width=0.48\textwidth]{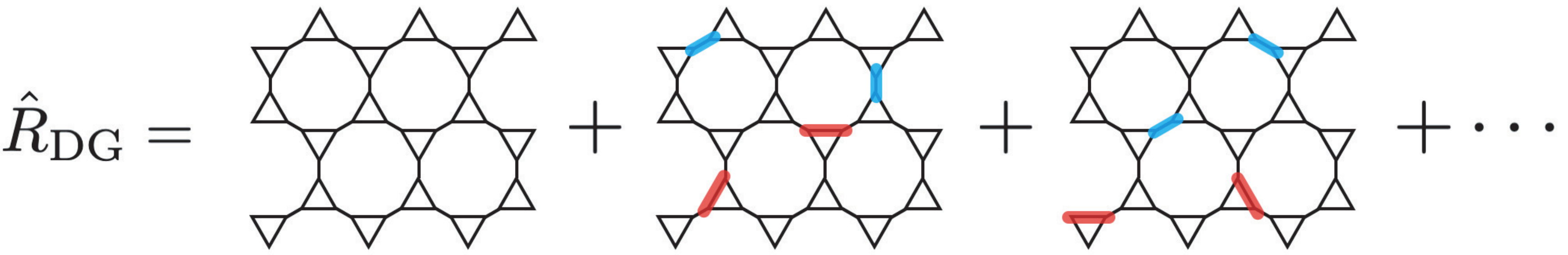},\nonumber
\end{align}
where the dimer on a bond $\langle ij \rangle$ stands for $\hat{\sigma}^{\gamma}_i \hat{\sigma}^{\gamma}_j$ depending on the bond, and different colors for the intra-bond and inter-bond. Note that one can introduce variational parameters in the DG operator, which determine the fugacities of the dimers\cite{HY19}, and optimize them to lower energy. Here, we introduce two independent fugacites such that the one on the intra-triangle bonds\,($c_1$) and another one on the inter-triangle bond\,($c_2$), i.e., $\hat{R}_{\rm DG}=\hat{R}_{\rm DG}(c_1,c_2)$. This operator does not spoil the symmetries and gauge structure of the ansatze\cite{HY19}. Then, we employ the state $|\psi_{\rm SG}(c_1,c_2)\rangle = \hat{Q}_{\rm LG} \hat{R}_{\rm DG}(c_1,c_2) |\Psi_{(111)}\rangle$ as our ansatz which can be regarded as the SG state\cite{HY19}. Here, the state $|\Psi_{(111)}\rangle \equiv |\Psi(\tan^{-1}\sqrt{2})\rangle$ is the the (111)-magnetic state where all spins point to $(111)$-direction.
In fact, one can apply the DG operator on the general product state $|\Psi(\theta)\rangle$ to have an additional variational parameter $\theta$. Instead, for brevity, we fix the initial product state as $|\Psi_{(111)}\rangle$ and optimize $c_1$ and $c_2$ for a given $\phi$. The obtained variational energy is presented in Fig.\,\ref{fig:energy}\,(a), of which the inset is the energy deviation from the exact one\,(see Appendix\,\ref{app:parameter} for details on the optimized parameters). As one can see, the DG operator drastically reduces the energy and provides reasonably good ansatz even around the transition point. Furthermore, the second derivative of the energy allows us to estimate the transition point correctly(see Appendix\,\ref{app:parameter}). Note that, in the non-Abelian phase, the variational energies are particularly good around $\phi=0.25\pi$. This is because $|\Psi_{(111)}\rangle$ is used as the initial state, which is optimal at $\phi=0.25\pi$. We thus believe that one can obtain even better ansatze\,($d E \sim O(10^{-4})$) throughout the non-Abelian phase by choosing the initial state $|\Psi(\theta)\rangle$ properly.

\section{minimally entangled states}

So far we have considered the ansatz only on the infinite system. Now we discuss the ansatz on the compact manifold\,(e.g., torus), 
%which requires more care, 
where the topological sectors allows us to distinguish the Abelian and non-Abelian phases. 
With periodic boundary conditions\,(PBC), one should also consider the so-called global flux measured by the flux operator $\hat{\Phi}_{\Gamma} = \prod_{i\in \Gamma} \hat{\sigma}_i^{\gamma_i}$ defined on a non-contractible closed path $\Gamma$\cite{Kitaev2006}. 
%This is because the KSM commutes with the global flux operator $\hat{\Phi}_{\Gamma} = \prod_{i\in \Gamma} \hat{\sigma}_i^{\gamma_i}$ where $\Gamma$ denotes a non-contractible closed path\cite{Kitaev2006}. 
Its eigenvalues $\pm1$ determine the topological sector, say even\,(odd) sector for $+1$\,($-1$). 
However, it turns out that our ansatze with PBC are not eigenstates of the global flux operators. 
To be more specific, let us consider ansatz on the torus\,(or an infinitely long cylinder) and then apply $\hat{\Phi}_y$ wrapping the inner tube of torus\,(say $y$-direction). 
%Using Eq.\,(3) in Ref.\,\cite{HY19}, 
One can verify that multiplying $\hat{\Phi}_y$ is equivalent to the gauge twisting along a closed path encircling the tube as illustrated below:
\begin{align}
  \includegraphics[width=0.48\textwidth]{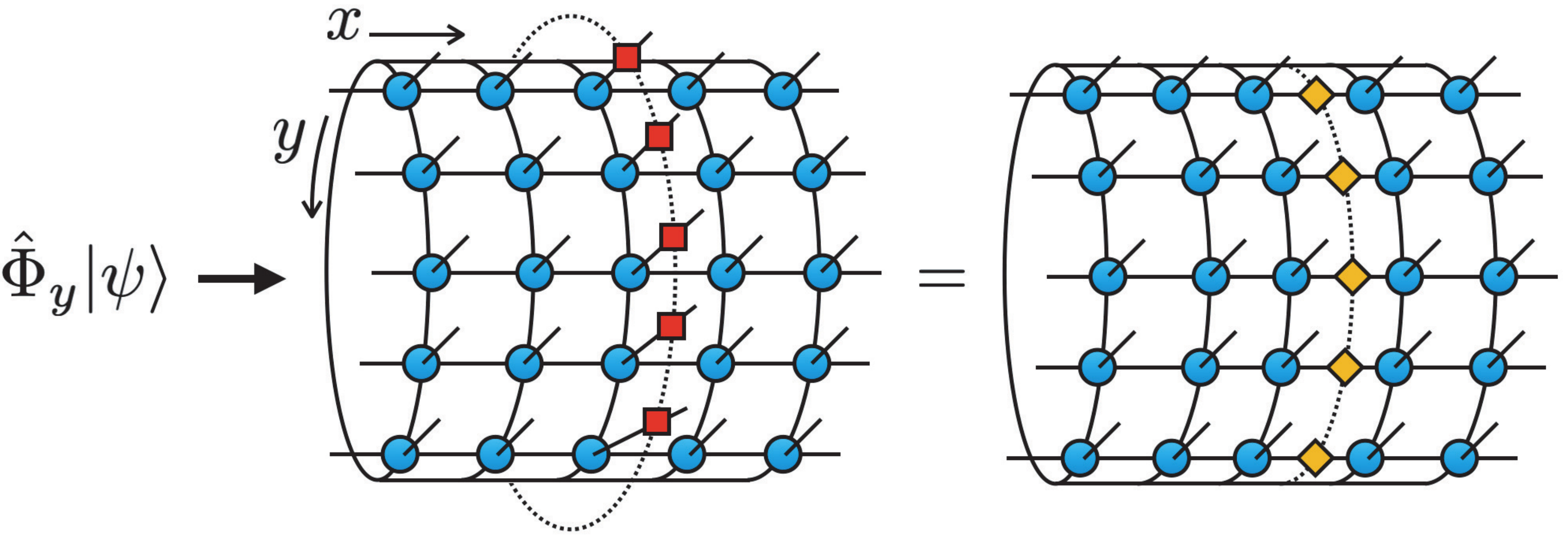},
  \label{eq:wilson_loop}
\end{align}
where the five-leg tensor is composed of six onsite  tensors in the unit-cell, and
\begin{align}
  \includegraphics[width=0.48\textwidth]{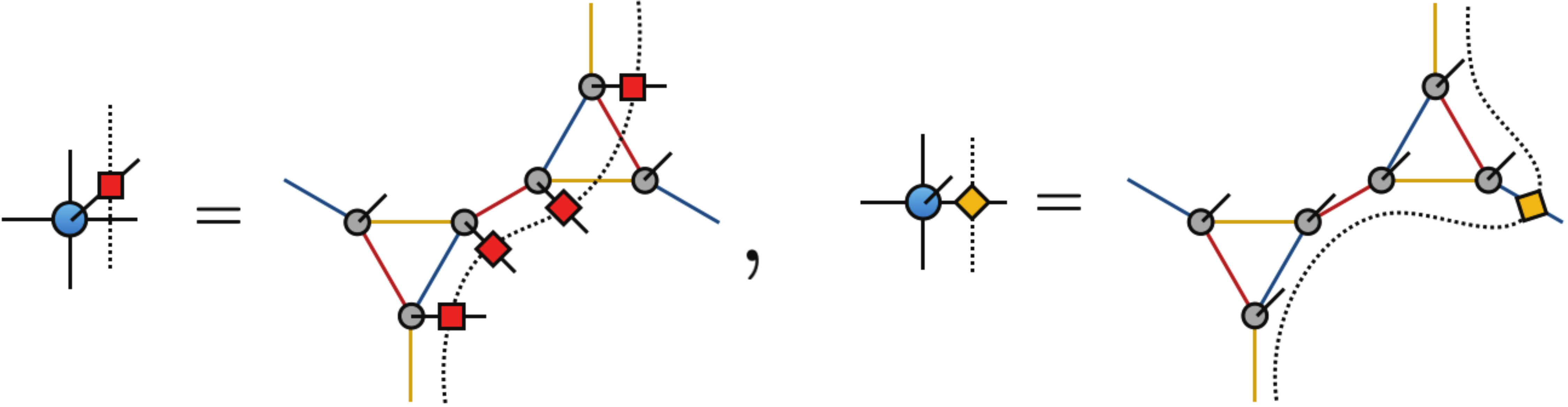}.\nonumber
\end{align}
Here, the red squares denote $\hat{\Phi}_y$ whereas the yellow one stands for the non-trivial element of the $Z_2$ invariant gauge group of our ansatz, $g=\hat{\sigma}^z$\cite{HY19}. The detailed derivation is presented in Appendix\,\ref{app:global}. Note that RHS of Eq.\,\eqref{eq:wilson_loop} is equivalent to the procedure of creating a vortex pair, moving one of those around the loop $\Gamma_y$, and then annihilating the pair. 
We define $G_y = \otimes_{i=1}^{L_y} g$ as the string of $g$ wrapping the inner tube where $L_y$ is the circumference in units of the unit-cell, i.e., the ring of yellow tensors in the right-hand side of Eq.\,\eqref{eq:wilson_loop}. Therefore, acting $\hat{\Phi}_y$ changes our ansatz $|\psi\rangle$\,(regardless of LG or SG) to a different state $|\psi_y \rangle$ that is the $G$-inserted $|\psi\rangle$. However, since the square of $\hat{\Phi}_y$ is identity, its eigenstates  are simply obtained by $|\psi_{\pm}\rangle = |\psi\rangle \pm |\psi_y\rangle$, and the subscript $\pm$ labels the global flux number, i.e., $\hat{\Phi}_y |\psi_{\pm}\rangle = \pm |\psi_{\pm}\rangle$. 
%Note that the insertion of $g$ on a single bond adds a minus sign to the configuration if a loop lies on the bond. 
In a similar way, one can set the simultaneous eigenstates of both $\hat{\Phi}_x$ and $\hat{\Phi}_y$, i.e., $|\psi_{(\pm,\pm)}\rangle$ living in one of four topological sectors specified by $(\hat{\Phi}_x,\hat{\Phi}_y)=(\pm1,\pm1)$.
Interestingly, in case of the LG ansatze, those in distinct sector are characterized by the parity of the number of non-contractible loop configurations in each direction\cite{Kitaev2003,Poiblanc12}. It can be easily seen that the action of $G_y$ gives a minus sign to all the configurations with odd number of non-contractible loops enclosing the hole of torus\,(say $x$-direction). 
%Consequently, the state $|\psi_{+\,(-)}\rangle$ consists of the configurations with only even\,(odd) number of loops enclosing the hole of torus.  Each sector is characterized by the parity of the number of loops encircling the hole and inner tube of torus. 
%as depicted in Fig.\,\ref{fig:sectors_torus}. 
Now, using those topologically degenerate ansatze, we construct the so-called minimally entangled states\,(MES)\,\cite{Zhang12}, e.g., $|\mathbb{I}\rangle = |\psi_{(+,+)}\rangle + |\psi_{(-,+)}\rangle$ and $|\mathfrak{m}\rangle = |\psi_{(+,-)}\rangle + |\psi_{(-,-)}\rangle$ characterized by each anyon\,($\mathbb{I}$: trivial, $\mathfrak{m}$: vortex) flux threading the inner tube of torus. In this basis, one can read off the quantum dimension\,($d_i$) of each anyon from the topological entanglement entropy\,(TEE), $\gamma_i = \log(D/d_i)$, where $i=\mathbb{I},\mathfrak{m},\cdots$. The total quantum dimension $D=\sqrt{\sum_i d_i^2}$\cite{Kitaev06a, Levin06, Dong08} is known to be four for the Abelian and non-Abelian KSL phases\cite{Yao10, Chung10}. 
To this end, we employ the bulk-edge correspondence in TPS\,\cite{Cirac11} to evaluate the entanglement spectrum\,(ES) and entanglement entropy\,(EE) on the infinitely long cylinder. Here, we impose PBC in the $y$-direction. 
\begin{figure}[!t]
  \includegraphics[width=0.5\textwidth]{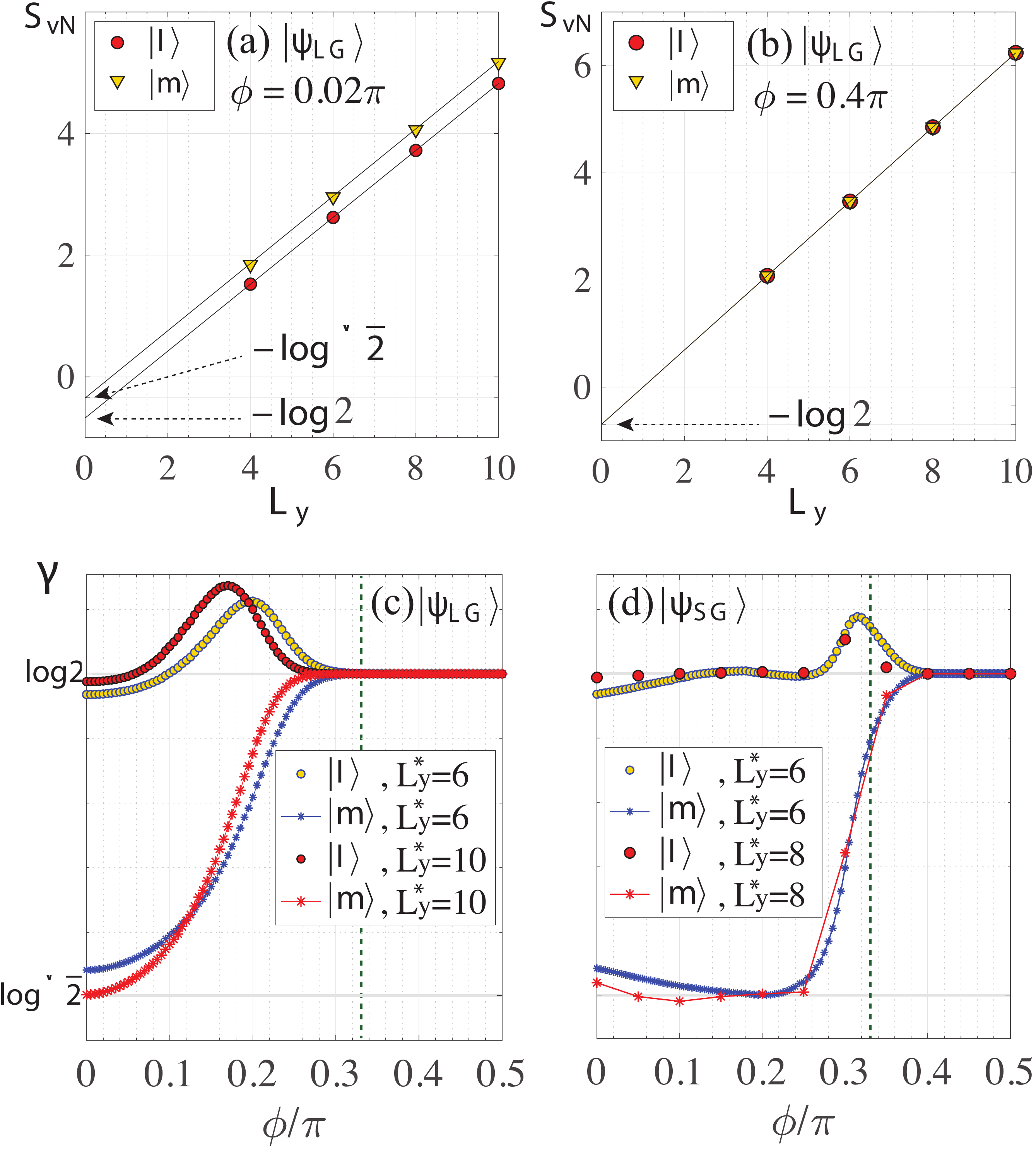}
  \caption{ The EE of $|\psi_{\rm LG}\rangle$ on the infinitely long cylnder at (a) $\phi=0.02\pi$ and (b) $\phi=0.4\pi$ as a function of $L_y$. Here, $|\mathbb{I}\rangle$ and $|\mathfrak{m}\rangle$ denote two degenerate MESs\,(see text), and the black solid lines are the fitting curves. Plots of the TEE $\gamma$ extracted from (c) the LG and (d) SG ansatze at each $\phi$, and the green dotted line denotes the critical point\,($\phi_c=\pi/3$), where $L_y^*$ denotes the largest circumference for fitting the data. For instance, the TEE $\gamma$ of $L_y^*=6$ is extracted by fitting the EE of $L_y = 4$ and $6$.
  %In $0\leq\phi<0.3\pi$, the TEE of $|\mathfrak{m}\rangle$ agrees with the one of the $\sigma$-anyon, i.e., $\gamma_{\mathfrak{m}} = \log \sqrt{2}$, in the Ising anyon model\cite{Kitaev2006, Nayak2008}. 
  }
  \label{fig:EE}
\end{figure}
Firstly, the results of TEE obtained from the LG and SG ansatze are presented in Fig.\,\ref{fig:EE}.  Here, (a) and (b) show the EEs in each sector obtained from $|\psi_{\rm LG}\rangle$ at $\phi=0.02\pi$ and $\phi=0.4\pi$, respectively, as a function of the circumference $L_y$. As expected from the geometry of TPS\cite{Verstraete06}, all EEs follow the area law\cite{Srednicki93}: $S = \alpha L_y - \gamma_i$ where $\alpha$ is a non-universal prefactor, and $\gamma_i$ is extracted by fitting the data with linear functions\,(black solid lines). At $\phi=0.02\pi$, we obtained $(\alpha, \gamma_i) = (0.5502, 0.6786)$ and $(0.5535, 0.3544)$ in each $\mathbb{I}$ and $\mathfrak{m}$ sector, respectively\,[Fig.\,\ref{fig:EE}\,(a)]. Those TEE are remarkably close to the TEE in the vacuum sector\,(i.e., $\log 2$) and the $\sigma$-anyon\,(vortex) sector\,(i.e., $\log \sqrt{2}$) of Ising anyon model\cite{Kitaev2006, Nayak2008}. On the other hand, at $\phi=0.4\pi$\,[Fig.\,\ref{fig:EE}\,(a)], both EEs almost perfectly fit to $(\alpha, \gamma_i) = (\log 2, \log 2)$, which is consistent with the one from the toric code\cite{Kitaev2003, Kitaev2006}. Similarly, we have extracted $\gamma_i$ at each $\phi$, and the results obtained from $|\psi_{\rm LG}\rangle$ and $|\psi_{\rm SG}\rangle$ are shown in Fig.\,\ref{fig:EE}\,(c) and (d), respectively. Those of $|\psi_{\rm LG}\rangle$ are in an excellent agreement with the ones of Ising anyon model around $\phi=0$ and with the ones of the toric code mostly in the Abelian phase\,[Fig.\,\ref{fig:EE}\,(c)]. Meanwhile, the SG ansatz gives almost consistent TEEs even in the non-Abelian phase agreeing with the ones of Ising anyon model and predicts the transition point rather correctly\,[Fig.\,\ref{fig:EE}\,(d)]. 
%Therefore, the vortex excitation in our ansatz might be identified with the non-Abelian $\sigma$-anyon in the non-Abelian phase while the vacuum with an Abelian anyon regardless of the phase.

%
\begin{figure}[!t]
  \includegraphics[width=0.5\textwidth]{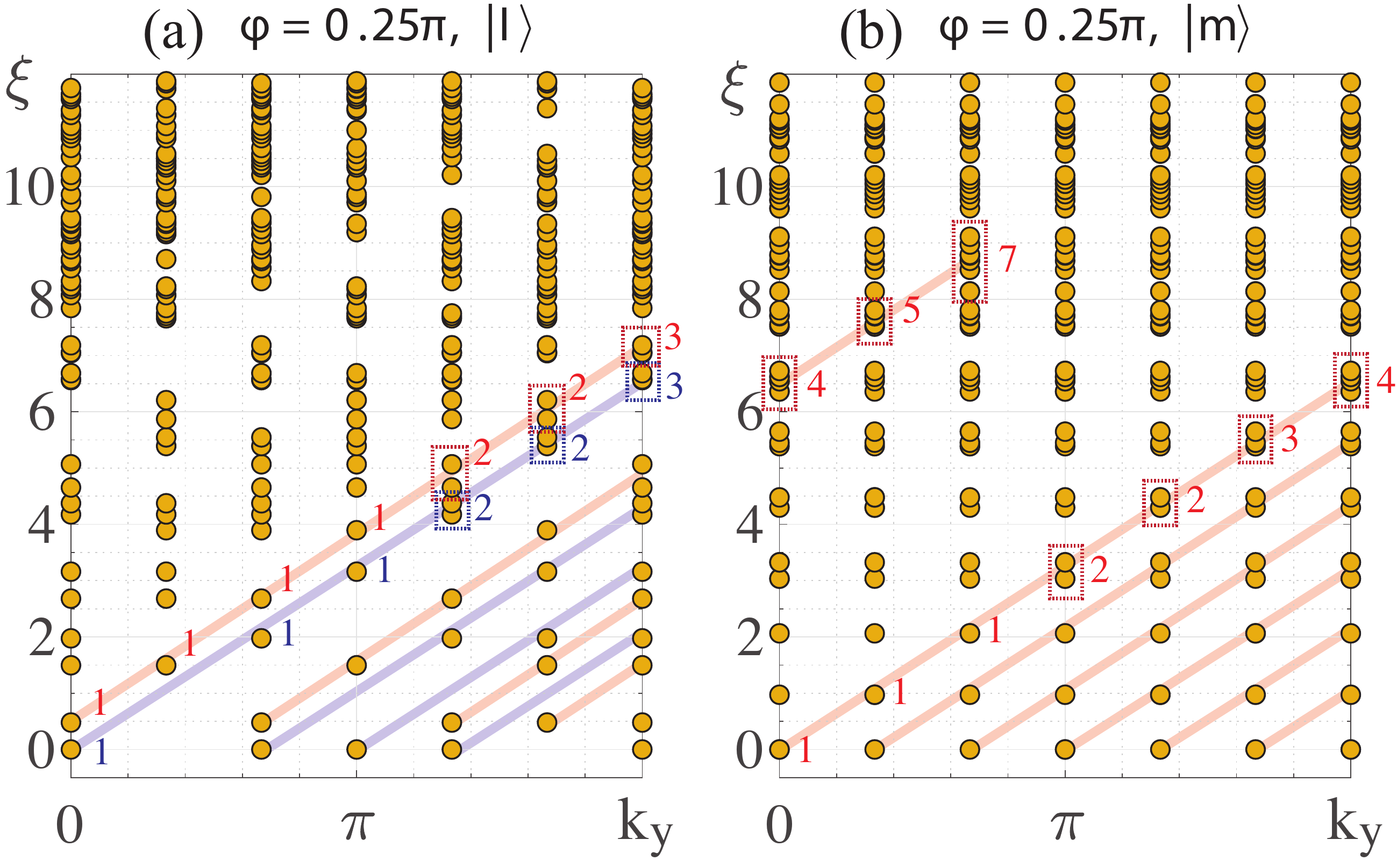}
  \caption{ The ES\,\cite{Li08} of two topologically degenerate ansatze $|\mathbb{I}\rangle$ and $|\mathfrak{m}\rangle$ with $L_y=6$ at $\phi=0.25\pi$. The level spacings and degeneracy patterns of chiral modes in (a) and (b) are consistent with three primary fields and their descendants in the Ising CFT\,(see text for details).  }
  \label{fig:ES_Ly6}
\end{figure}
%d

Furthermore, the identification of the topological excitations becomes even clearer from characteristic structures in the ES\cite{Li08}. %which are shown in Fig.\,\ref{fig:ES_Ly6}. 
%Here, we have used the translational symmetry along the $y$-direction to block-diagonalize the reduced density matrices obtained from the SG ansatze\,($L_y=6$) such that the ESs are labeled by the momentum $k_y$. 
Figure\,\ref{fig:ES_Ly6}\,(a) and (b) present the ESs  of $|\mathbb{I}\rangle$ and $|\mathfrak{m}\rangle$  obtained by the SG ansatz at $\phi=0.25\pi$. Here, the circumference is $L_y=6$, and the horizontal axis $k_y$ denotes the momentum. 
There are four branches of two distinct chiral modes in the $\mathbb{I}$-sector, which linearly disperse in one direction. Those are highlighted by the red and blue solid lines in Fig.\,\ref{fig:ES_Ly6}\,(a). 
Assuming the close ESs\,(dashed boxes) as degenerate levels, the degeneracy pattern is consistent with the ones of the primary fields ${\bf 1}$\,(blue) and $\psi$\,(red) and their descendants in the Ising CFT\cite{Friedan84, Henkel99}, respectively. On the other hand, in the $\mathfrak{m}$-sector, we find six branches of a single chiral mode of which the degeneracy counting obeys $\{1,1,1,2,2,3,4,5,7,\cdots\}$\cite{Henkel99}, i.e., the characteristic of the primary field $\sigma$ and its descendants in the Ising CFT. In addition, the level spacings in the low-lying spectrum are in excellent agreement with the exact ones\,(see Appendix\,\ref{app:spectrum}). From our MES setup, the state $|\mathfrak{m}\rangle$ is expected to accommodate the vortex at each boundary, and thus the vortex is identified with the $\sigma$-anyon exhibiting non-Abelian braiding statistics\cite{Henkel99, Nayak2008}. 
%We also found similar ESs in $0<\phi\leq 0.25\pi$, and it directly indicates that the SG ansatz in the non-Abelian phase hosts the non-Abelian anyons correctly.
%As shown in Fig.\,\ref{fig:ES_Ly6}\,(c) and (d), in the Abelian phase, the ESs are completely gapped from the lowest ones in both sectors. The dispersionless lowest levels can be interpreted as the formation of dimer between spins on the inter-triangle bonds. Consequently, the boundary spins do not communicate with each other, and hence those contribute equally to the entanglement.

%
\section{conclusion}

In this Letter, we show that the Abelian and non-Abelian CSL ground states of the KSM are well represented by the LG and SG states. 
In particualr, at both limits $\phi = 0$ and $\pi/2$, the LG states become exact. Further, the gap closing at $\phi=0$ is understood by mapping the norm of ansatz into the partition function of the critical LG model. In addition, the fate of long-ranged loops is found to determine the Abelianess and non-Abelianess of CSL. By constructing the MES and measuring its TEE, we directly show that our ansatze host indeed the non-Abelian vortex with the quantum dimension $d_{\mathfrak{m}}=\sqrt{2}$. On the other hand, it becomes trivial, i.e., $d_{\mathfrak{m}}=1$, as the ansatz enters into the Abelian phase. We also identify the chiral edge modes in the non-Abelian phase with the Ising CFT, not the SU(2)$_2$ Wess-Zumino-Witten theory conjectured in Ref.\,\cite{Yao07} by exploiting the level spacing and their degeneracy patterns\cite{Francesco12,Chen18}. We believe that the LG ansatze are the simplest CSLs in a compact representation, and therefore it could provide a platform bridging the quantum loop models\cite{Fendley2008} with the Abelian and non-Abelian topological states. It is also worth noting that our ansatze are the rigorous example explicitly revealing that general TPSs can represent the chiral gapped states. In the case of the fermionic Gaussian TPS, there exists a no-go theorem\cite{Dubil15} prohibiting the chiral Gaussian TPS to be gapped. However, it was not so clear whether the theorem applies to generic TPSs or not\cite{Poiblanc15, Shuo15, Poiblanc16, Poiblanc17, Chen18}. 
We believe that our ansatze are the counter evidence against the generalization of the theorem. In technical aspects, two independent optimization schemes, which can be combined together, are introduced for the LG and SG ansatz. Those can be employed to study the anisotropic Kitaev model on the honeycomb lattice and its extensions which are relevant to Kitaev materials such as $\alpha$-RuCl$_3$\cite{Jackeli2009, Chaloupka2010, Singh2010, Kimchi11, Gohlke2018, Banerjee2016, Banerjee2018}. Also, the variational ansatze are of interest of the deformed topological wavefunction\cite{Xu19}.

\begin{acknowledgements}
	The computation in the present work was executed on computers at the Supercomputer Center, ISSP, University of Tokyo, and also on K-computer (project-ID: hp190196). N.K.'s work is funded by MEXT KAKENHI No.19H01809. H.-Y.L. was supported by MEXT as ``Exploratory Challenge on Post-K computer"\,(Frontiers of Basic Science: Challenging the Limits). T.O. was  supported by JSPS KAKENHI No.15K17701 and 19K03740. R.K. was supported by MEXT as "Priority Issue on Post-K computer" (Creation of New Functional Devices and High-Performance Materials to Support Next-Generation Industries).	
\end{acknowledgements}

\onecolumngrid
\appendix

\section{ Norm of the zeroth ansatz }
\label{app:norm}

In this section, we show that the norm of the loop gas\,(LG) ansatz $|\psi_0(\theta)\rangle = \hat{Q}_{\rm LG} |\Psi(\theta)\rangle$ can be exactly mapped into the partition function of the $O(1)$ loop gas model on the honeycomb lattice. To this end, we first note that the LG operator is an hermitian projector: $\hat{Q}_{\rm LG}^{\dagger} = \hat{Q}_{\rm LG}$ and $(\hat{Q}_{\rm LG})^2 = N_{\Gamma} \hat{Q}_{\rm LG} $ where $N_{\Gamma}$ is the total number of loop configurations on the star lattice. The LG operator is efficiently represented by the tensor product operator\cite{HY19}, i.e., $\hat{Q}_{\rm LG} = {\rm tTr} \prod_{\alpha} Q_{i_{\alpha} j_{\alpha} k_{\alpha}}^{ss'}|s\rangle\langle s'|$ where tTr stands for the tensor trace, $\alpha$ labels the site index, the building block tensor

\begin{align}
	Q_{i j k}^{s s'} = \tau_{i j k } [(\hat{\sigma}^x)^{1-i} (\hat{\sigma}^y)^{1-j} (\hat{\sigma}^z)^{1-k}]_{ss'},\quad\quad
	\tau_{i j k} = \begin{cases}
		-i \quad &{\rm if}\quad i+j+k=0\\
		1 \quad &{\rm if}\quad i+j+k=2
	\end{cases},
	\label{eq:lg_operator}
\end{align}
and the indices of physical and virtual legs are $s,s'=0,1$ and $i,j,k=0,1$, respectively. Also, the virtual legs $i,j$ and $k$ lie on the $x,y$ and $z$ bonds in the model\,(see Fig.\,1\,(a) in the main text), respectively. Since the LG operator is obtained by summing over all possible loop operators\,(that is, product of $\hat{\sigma}^x$, $\hat{\sigma}^y$ and $\hat{\sigma}^z$ along the loops), it is straightforward to verify $(\hat{Q}_{\rm LG})^2 = N_{\Gamma} \hat{Q}_{\rm LG}$ using the manipulation rules of loop defined in Ref.\,\cite{HY19}. One can also easily show its hermiticity using the $Q$-tensor in Eq.\,\eqref{eq:lg_operator}. Now, let us compute the norm of the LG ansatz:

\begin{align}
	\langle \psi_0(\theta) | \psi_0(\theta) \rangle 
	= \langle \Psi(\theta) | \hat{Q}_{\rm LG}^{\dagger} \hat{Q}_{\rm LG} | \Psi(\theta) \rangle = N_{\Gamma} \langle \Psi(\theta) | \hat{Q}_{\rm LG} | \Psi(\theta) \rangle = N_{\Gamma} \sum_{G\in \Gamma} x_G(\theta),
\end{align}
where we used the hermiticity and idempotence\,(up to overall $N_{\Gamma}$), and $G$ denotes each loop configuration, $x_G(\theta) = \langle \Psi(\theta) | \hat{Q}_G | \Psi(\theta)\rangle$ is the weight of a loop operator $\hat{Q}_G$. With the definition of $|\Psi(\theta)\rangle$ and Eq.\,(2) in the main text, one can easily verity $x_G(\theta) = (\sin \theta)^{l_G^t} (\cos \theta/\sqrt{2})^{l_G^d}$ where $l_G^t$ and $l_G^d$ are the total lengths of partial loops along the triangle and dodecagon plaquettes, respectively, in the configuration $G$. Now, let us consider local loop configurations on a single triangle plaquette\cite{Batchelor1998}. There are eight configurations on the single triangle plaquette which are depicted in the left hand side of graphical equations below:

\begin{align}
  \includegraphics[width=0.6\textwidth]{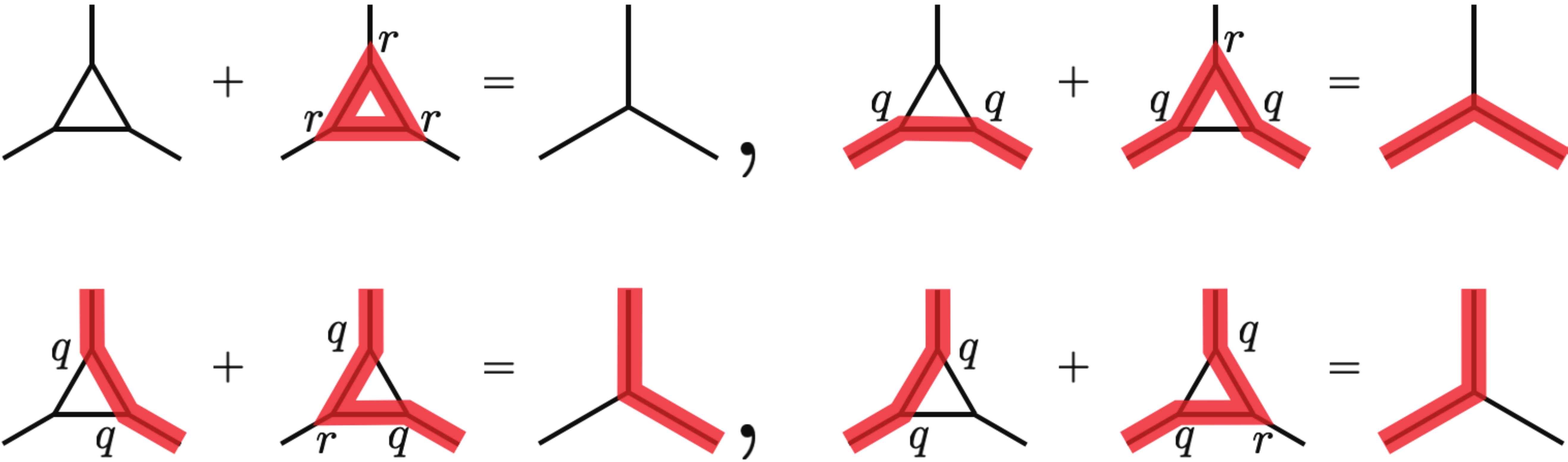},
  \label{eq:mapping}
\end{align}
where $r = \cos\theta$ and $q = \sin\theta /\sqrt{2}$ as defined in the main text. As shown equation above, summing two of them can be regarded as a configuration on the honeycomb lattice by treating the triangle plaquette as a single site. Then, our task falls into the calculation of the partition function of the $O(1)$ LG model on the honeycomb lattice, where the weights of hole and loop per site are $1+r^3$ and $q^2(1+r)$, respectively, and it is simply given by

\begin{align}
	\langle \psi_0(\theta) | \psi_0(\theta) \rangle 
	= N_{\Gamma} \sum_{G' \in \Gamma'} (1+r^3)^{n-n_{G'}} [q^2(1+r)]^{n_{G'}}
	= N_{\Gamma} (1+r^3)^n \sum_{G' \in \Gamma'} \left( \frac{q^2(1+r)}{1+r^3}\right)^{n_{G'}}.
\end{align}
Here, $\Gamma'$ denotes all possible loop configurations on the honeycomb lattice, and $n$ is the total number of sites on the honeycomb lattice while $n_{G'}$ stands for the total length of loops in a configuration $G'$. The RHS is identical to the partition function of the $O(1)$ LG model on the honeycomb lattice, i.e., $Z_{O(1)}(x)$, with the loop fugacity $x = q^2/(1-r+r^2)$, of which the critical point is $x_c = 1/\sqrt{3}$\cite{Nienhuis1982}. Consequently, the norm of the LG ansatz maps to $Z_{O(1)}(q^2/(1-r+r^2))$ which becomes critical at $q_c = \sqrt{(1-r+r^2)/\sqrt{3}}$ for a given $r$.

\section{ optimal variational parameters in the first order ansatz }
\label{app:parameter}
\begin{figure}[!b]
  \includegraphics[width=0.8\textwidth]{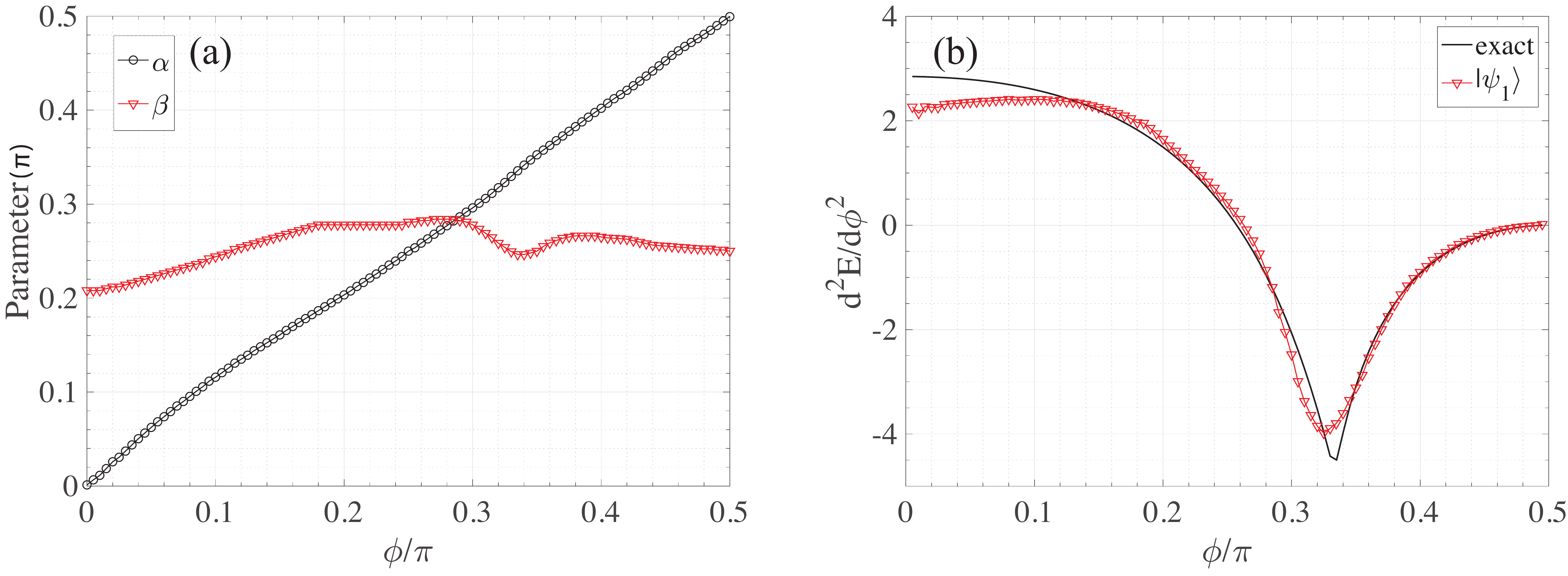}
  \caption{ (a) The optimal variational parameters $\alpha$ and $\beta$ in the first order ansatz $|\psi_1(\alpha,\beta)\rangle$ as a function of $\phi$. (b) The second derivative of the energy of the first order ansatz in terms of $\phi$. }
  \label{fig:parameters}
\end{figure}

In this section, we present optimal variational parameters in the first order ansatz $|\psi_1(c_1,c_2)\rangle = \hat{Q}_{\rm LG} \hat{R}_{\rm DG}(c_1, c_2)|\Psi_{111}\rangle $, where $|\Psi_{111}\rangle$ is the product state of local magnetic (111)-state. In a similar way to the LG operator, the dimer gas\,(DG) operator\cite{HY19} $\hat{R}_{\rm DG}(c_1, c_2)$ is also efficiently represented in the tensor network and by the following building block tensor

\begin{align}
	R_{i j k}^{ss'} = \zeta_{i j k} \left[ (\hat{\sigma}^x)^i (\hat{\sigma}^y)^j (\hat{\sigma}^z)^k \right]_{ss'},\quad\quad
	\zeta_{i j k} = \begin{cases}
		1 \quad &{\rm if}\quad i+j+k=0\\
		c_1/c_2 \quad &{\rm if}\quad i+j+k=1
	\end{cases},
\end{align}
where we assign $c_1\,(c_2)$ if the non-zero element comes from the intra-triangle\,(inter-triangle) bond. The dimension and direction of the virtual indices $i,j,$ and $k$ are the same as the ones of the $Q$-tensor in Eq.\,\eqref{eq:lg_operator}. To be more specific, on the site 8 in Fig.\,1\,(a) in the main text on which the inter-triangle bond is the $y$-bond, we put the building block tensor with 

\begin{align}
	\zeta_{i j k}^{\rm site\,8} = \begin{cases}
		1 \quad &{\rm if}\quad i+j+k=0\\
		c_1 \quad &{\rm if}\quad j=0 \quad {\rm and}\quad k+i=1\\
		c_2 \quad &{\rm if}\quad j=1 \quad {\rm and}\quad k+i=0
	\end{cases}.
\end{align}
Note that the variational parameters $c_1$ and $c_2$ determine the fugacity of the dimer on the intra-triangle bond and inter-triangle bond, respectively, whereas the fugacity of hole is set to unity. Let us reparametrize the variational parameters as follows: $\hat{R}_{\rm DG}(c_1, c_2) \rightarrow \hat{R}_{\rm DG}(\alpha, \beta)$ with

\begin{align}
	\zeta_{i j k} = \begin{cases}
		\cos \beta \quad &{\rm if}\quad i+j+k=0\\
		\sin \beta \sqrt{\cos \alpha} /\sin \beta \sqrt{\sin \alpha} \quad &{\rm if}\quad i+j+k=1 
	\end{cases}.
\end{align}
Now, one can vary two variational parameters $\alpha$ and $\beta$ to minimize the energy expectation value of the Kitaev model on the star lattice. Using the corner transfer matrix renormalization group method, we measured the expectation values and then found the optimal $\alpha$ and $\beta$ at a given $\phi$\,(see the main text) which are presented in Fig.\,\ref{fig:parameters}\,(a). The energy expectation values shown in Fig.\,2(a) in the main text are obtained with the optimal parameters presented in Fig.\,\ref{fig:parameters}\,(a). Interestingly, the optimal $\alpha$, which determines the relative weight between dimers on the intra-triangle and inter-triangle bonds, increases linearly with the model parameter $\phi$, which determines the relative strength between the exchange couplings on the intra-triangle\,($J$) and inter-triangle\,($J'$) bonds. Meanwhile, the parameter $\beta$, which determines the relative weight between the hole and dimers, is optimized in $0.2\pi < \beta < 0.3\pi$. Note that one can also introduce complex fugacities, which lead to two more variational parameters, i.e., $(c_1, c_2) \rightarrow (c_1 e^{i\eta_1}, c_2 e^{i\eta_2})$. However, we found that real fugacities always give the lowest energy throughout the model parameter $\phi$. Figure\,\ref{fig:parameters}\,(b) shows the second derivative of the energy expectation value of $|\psi_1\rangle$ in terms of $\phi$. Here, the transition point expected from $|\psi_1\rangle$ is a bit different from the exact one $\phi_c = \pi/3$. However, considering the fact that the curve is obtained by numerical differenciation twice, its accuracy and smoothness are quite remarkable.

\section{global flux}
\label{app:global}
\begin{figure}[!t]
  \includegraphics[width=0.98\textwidth]{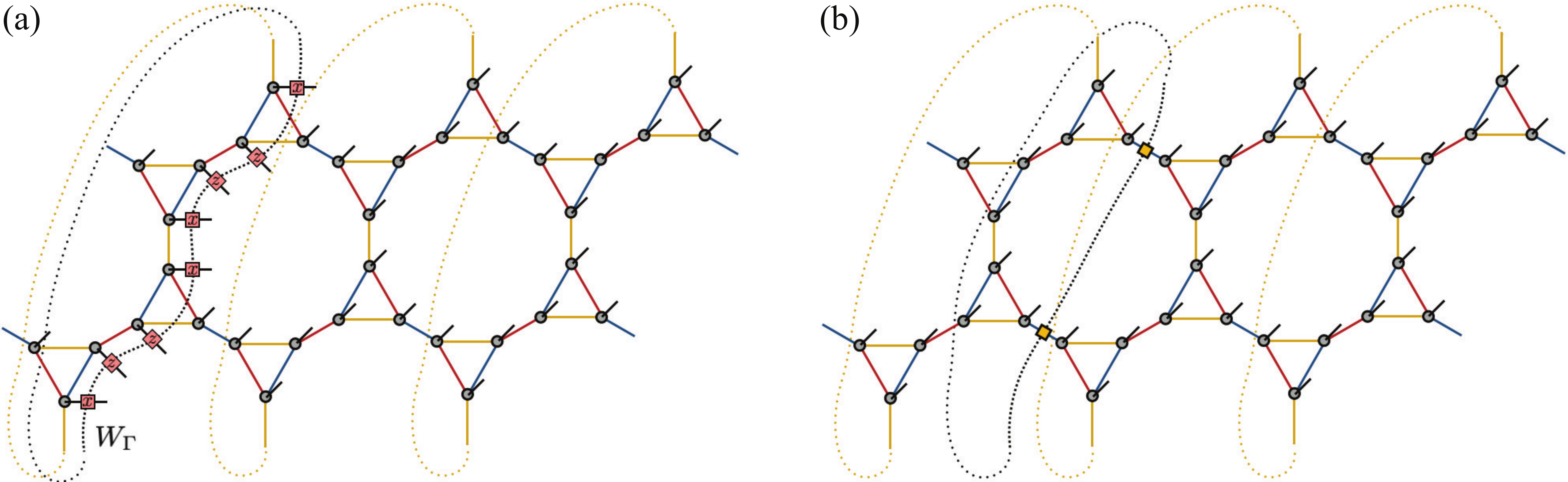}
  \caption{ Schematic figure of the action of \,(a) the global flux operator $W_{\Gamma}$ on the tensor product states and (b) the string of the non-trivial element\,(yellow square) of the invariant gauge group of the LG operator along the $y$-direction. }
  \label{fig:wilson_loop}
\end{figure}

Here, we show that the LG and SG ansatze are not the eigenstates of the global flux operators, i.e., the Wilson loop operator $W_{\Gamma} = \prod_{i\in \Gamma} \sigma_i^{\alpha_i}$ along a non-contractible loop $\Gamma$ on a compact manifold where $\alpha = x, y$ and $z$ depending on the site $i$\cite{Kitaev2006}. To this end, we first note that, as shown in Ref.\,\cite{HY19}, the multiplication of the Pauli matrices on the physical leg of local tensor of the LG operator is identical to the multiplication of the matrix 

\begin{align}
	v = \begin{pmatrix}
		0 && i \\ 1 && 0
	\end{pmatrix},
\end{align}
and its conjugate on two virtual legs:

\begin{align}
	\sigma^x_{s s''} Q_{ijk}^{s'' s'} = v_{jj'} v_{kk'}^* Q_{ij'k'}^{ss'},\quad\quad
 	\sigma^y_{s s''} Q_{ijk}^{s'' s'} = v_{kk'} v_{ii'}^* Q_{i'jk'}^{ss'},\quad\quad
	\sigma^z_{s s''} Q_{ijk}^{s'' s'} = v_{ii'} v_{jj'}^* Q_{ij'k'}^{s s'}.
	\label{eq:sigma}
\end{align}
Here, the tensor $Q_{ijk}^{s s'}$ denotes the local tensor of the LG operator, and repeated indices are implicitly summed over. Above relation can be simply described by the following graphical representation:

\begin{align}
  \includegraphics[width=0.8\textwidth]{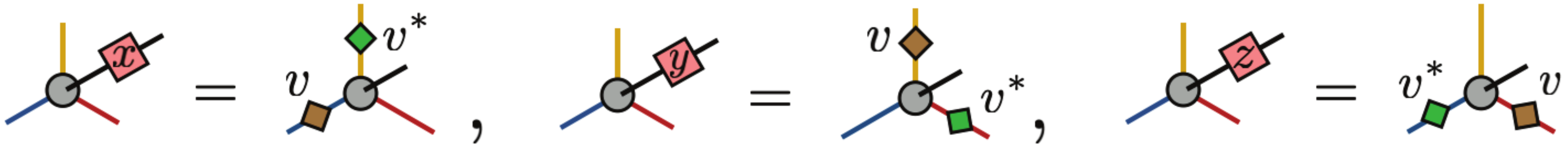},
  \label{eq:mapping}
\end{align}
where the gray circle denotes the $Q$-tensor, black solid line stands for the physical leg\,(index $s$) and red, blue and yellow solid lines are virtual legs\,(indices $i,j$ and $k$) on the $x, y$ and $z$ bonds of the Kitaev model. Also, the red square attached on the physical leg denotes the Pauli matrix. Note that a physical leg $s'$ is omitted in the graphical representation for simplicity. In what follows, using the above relation, we show how the LG and SG ansatze react on the action of the global flux operator $W_{\Gamma}$ and how to construct the eigenstates of $W_{\Gamma}$ using them. Let us consider the periodic boundary condition along the $y$-direction as defined in the main text and then act the operator $W_{\Gamma}$ along the $y$-direction as depicted in Fig.\,\ref{fig:wilson_loop}\,(a). Now, using the relation in Eq.\,\eqref{eq:sigma}, one can show that the ring of the tensor network, where the operator $W_{\Gamma}$ is applied, has the following equalities,

\begin{align}
  \includegraphics[width=0.98\textwidth]{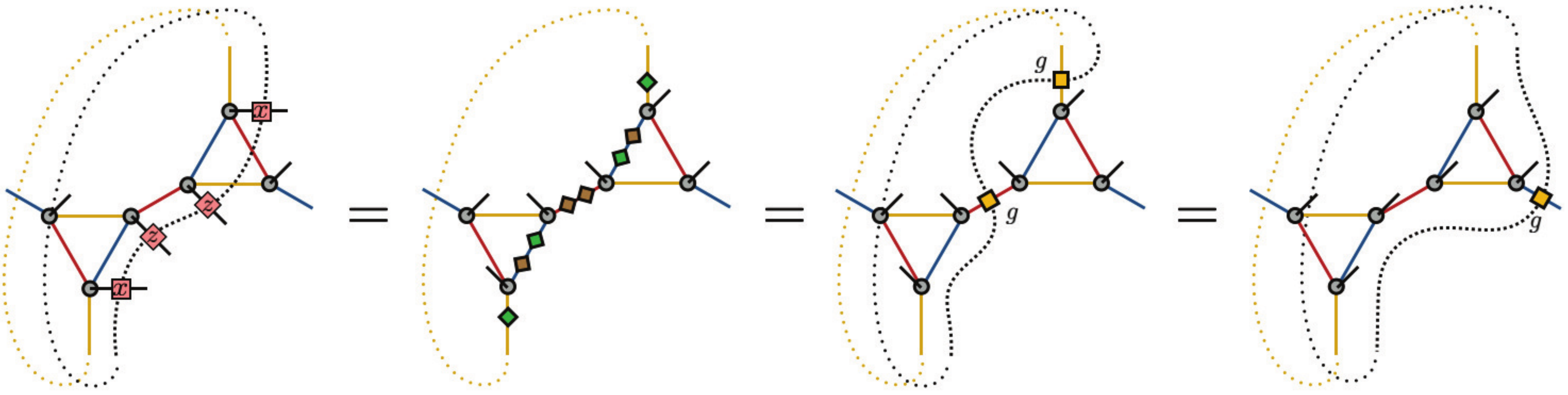},
  \label{eq:mapping}
\end{align}
where the matrix $g = \sigma^z$ is the non-trivial element of the $Z_2$ invariant gauge group of the LG operator\cite{HY19}. In the first equality, the relation in Eq.\,\eqref{eq:sigma} is applied, and we use relations $v v^{\dagger} = 1$ and $v v^{T}  = -\sigma^z$. Finally, in the last equality, the invariant gauge symmetry is used, i.e., $g_{i i'} g_{j j'} g_{k k'}Q_{i'j'k'}^{ss'} = Q_{ijk}^{ss'}$. Therefore, applying the operator $W_{\Gamma}$ on our ansatze results in a different tensor network where a string of $g$, $G = \prod_{i=1}^{L_y} g$, along the $y$-direction is inserted in the original state as illustrated in Fig.\,\ref{fig:wilson_loop}\,(b). One can easily notice that such a $G$ on the non-contractible loop can not be eliminated by a gauge transformation, and therefore the resulting state is not identical to the original state, i.e. Fig.\,\ref{fig:wilson_loop}\,(a) = Fig.\,\ref{fig:wilson_loop}\,(b). That is, our ansatze are not the eigenstate of the global flux operator. However, since $(W_{\Gamma}) = 1$ and $g^2=1$, it is easy to construct the eigenstate of $W_\Gamma$ using our ansatz, i.e., $|\psi_{\pm}\rangle = |\psi\rangle \pm |\psi_{G}\rangle$ where $|\psi\rangle$ denotes the LG or SG ansatz and $|\psi_{G}\rangle$ the $G$-inserted $|\psi\rangle$ along the $y$-direction as shown in Fig.\,\ref{fig:wilson_loop}\,(b). Then, the state $|\psi_\pm\rangle$ is the eigenstate of $W_\Gamma$ with the eigenvalue or global flux number $\pm1$.

\section{ entanglement spectrums and conformal towers }
\label{app:spectrum}
\begin{figure}[!t]
  \includegraphics[width=0.9\textwidth]{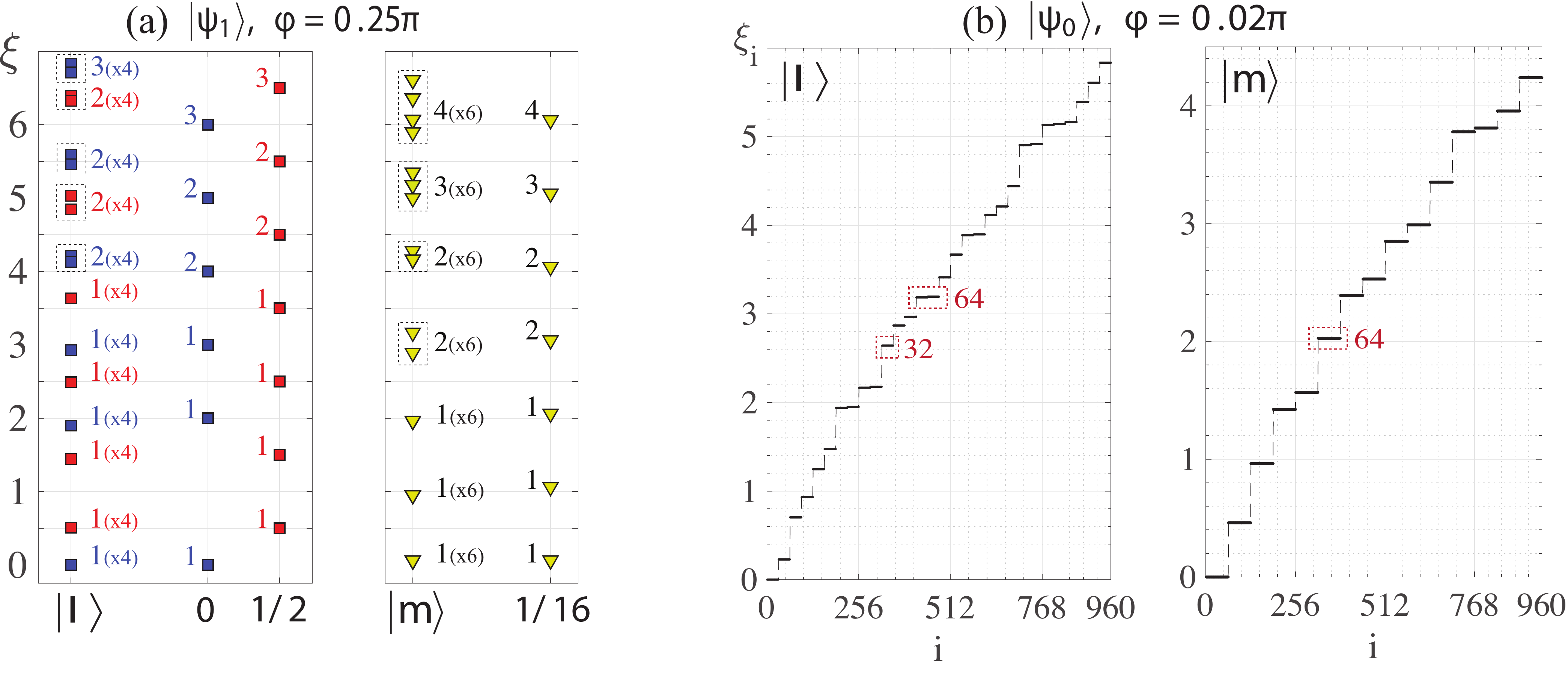}
  \caption{ The entanglement spectrums of (a) the first order ansatz\,($L_y=6$) at $\phi=\pi/4$ and (b) the zeroth order ansatz\,($L_y=12$) at $\phi=0.02\pi$. Here, $|\mathbb{I}\rangle$ and $|\mathfrak{m}\rangle$ denote the identity and Ising anyon sector, respectively. In (a), the Virasoro towers of the primary fields with the conformal weights $\Delta = 0, 1/2$ and $1/16$ of the Ising conformal field theory. The degeneracy is specified next to each level.}
  \label{fig:conformal_tower}
\end{figure}

In the main text, the entanglement spectrums are presented as a function of the momentum $k_y$ in each topological sector, and their degeneracy patterns are discussed. Here, we directly compare the entanglement spectrums and the Virasoro towers of the Ising conformal field theory. In Fig.\,\ref{fig:conformal_tower}\,(a), we compare the entanglement spectrums of the string gas\,(SG) ansatz $|\psi_1\rangle$ optimized at $\phi=0.25\pi$ with the Virasoro characters of the primary operators having the conformal weights $\Delta = 0, 1/2$ and $1/16$\cite{Henkel99}. As one can see, the spectrums in the sector $|\mathbb{I}\rangle$\,(left panel) can be regarded as the sum of two Virasoro towers of $\Delta = 0$ and $1/2$, while the ones in the sector $|\mathfrak{m}\rangle$\,(right panel) match with the Virasoro tower of $\Delta = 1/16$. Furthermore, their spacings and degeneracy patterns are in excellent agreement with the exact ones up to the seventh level. In Fig.\,\ref{fig:conformal_tower}\,(b), the entanglement spectrums of the loop gas ansatz $|\psi_0\rangle$ optimized at $\phi=0.02\pi$ are presented. However, we could not identify their characteristics and relationship to the Ising conformal field theory, though the ansatz provides very accurate variational energy and entanglement entropy as shown in the main text. 

\twocolumngrid

\bibliographystyle{apsrev}
\bibliography{reference.bib}
\end{document}